\documentclass[12pt]{article}
\usepackage{epsfig}
\usepackage{amsmath,amsfonts,amssymb}

\newcommand{\be}{\begin{equation}}
\newcommand{\ee}{\end{equation}}
\newcommand{\bem}{\begin{displaymath}}
\newcommand{\eem}{\end{displaymath}}
\newcommand{\ba}{\begin{eqnarray}}
\newcommand{\se}{\setcounter{equation}{0}}
\newcommand{\ea}{\end{eqnarray}}
\newcommand{\re}[1]{(\ref{#1})}

\newcommand{\1}{^{-1}}

\newcommand{\dg}{^{\dagger}}

\newcommand{\e}{\mbox{e}}

\newcommand{\f}{{\mbox{\scriptsize f}}}

\newcommand{\bG}{\bar{G}} 
 
\newcommand{\ga}{\gamma_5}
\newcommand{\h}{\frac{1}{2}}
\newcommand{\Id}{\mbox{1\hspace{-1.05mm}l}}   

\newcommand{\la}{\lambda}

\newcommand{\mo}[1]{^{\mbox{\scriptsize #1}}}
\newcommand{\N}{{\cal N}}

\newcommand{\bP}{\bar{P}}

\newcommand{\bS}{\bar{S}} 
 
\newcommand{\Tr}{\mbox{Tr}} 
\newcommand{\bu}{\bar{u}}

\newcommand{\vp}{\varphi} 
\newcommand{\bw}{\bar{w}}

\begin{document}
 
\hfill {\sc HU-EP}-05/17
 
\vspace*{1cm}
 
\begin{center}
 
{\Large \bf CP and T violation in non-perturbative\\chiral gauge theories}
 
\vspace*{0.9cm}
 
{\bf Werner Kerler}
 
\vspace*{0.3cm}
 
{\sl Institut f\"ur Physik, Humboldt-Universit\"at, D-12489 Berlin,
Germany}
 
\end{center}

\vspace*{1cm} 

\begin{abstract}
We give a completely general derivation revealing the precise origin and
the quantitative effects of CP and T violations in chiral gauge theories 
on the lattice. 

\end{abstract}

\section{Introduction}

Imposing the Ginsparg-Wilson (GW) relation \cite{gi82} $\{\ga,D\}=D\ga D$
on the Dirac operator and considering a special form of the chiral projections 
Hasenfratz \cite{ha02} has pointed out that the usual CP symmetry does not 
hold on the lattice. In our notation, writing the chiral projections as 
$P_-=\h(\Id-\ga G)$, $\;\bP_+=\h(\Id+\bG\ga)$, this form is given by 
\be
G=\big(\Id-sD\big)/\N,\quad\bG=\big(\Id-(1-s)D\big)/\N,
\label{Gs}
\ee
with a real parameter $s$ and $\N=\sqrt{\Id-s(1-s)DD\dg}$. CP 
violation then has been traced back to the singularity of \re{Gs} for $s=\h$ 
which does not allow to accomodate the interchange of $s$ and $1-s$ under
the transformation in a symmetric way. This has been extended to some more 
general Dirac operators in Ref.~\cite{fu02}, which in \re{Gs} amounts to a 
replacement of $D$ by $DF$ with a Hermitian function $F(DD\dg)$. 

It should be noted here that the chiral projections for GW fermions 
implicit in Ref.~\cite{na93} and used in Ref.~\cite{lu98} in the present 
notation correspond to the choice $G=\Id-D$, $\bG=\Id$. We also note that 
the generalized chiral symmetry proposed in Ref.~\cite{lu98a} can be more 
generally formulated in terms of $G$ and $\bG$ as $\e^{i\varepsilon\bG\ga}D
\e^{i\varepsilon\ga G}=D$ (the non-unitary choice $1-\h D$ in the GW case 
in Ref.~\cite{lu98a}, however, being not a legitimate one for $G$ and $\bG$). 

Recently Hasenfratz and Bissegger \cite{ha05}, again using the special form 
of Ref.~\cite{ha02} for the chiral projections of GW fermions, have also 
considered T transformations. They find violation of the symmetry quite 
similarly as in the CP case. In this context they also point out that CPT 
symmetry remains intact.

Obviously the form of the chiral projections based on \re{Gs} represents a 
rather special case. Thus firstly the question arises whether the observed 
symmetry violations really persist in general. Secondly instead of only 
tracing the violations back to a parameter singularity to reveal the precise 
reason for them is preferable. Thirdly then to get quantitative hold of the 
violations is desirable.

In a more general approach \cite{ke03} it has been shown that the symmetric 
situation of contiuum theory in the CP case is not admitted due to certain
operator properties. The operators $G$, $\bG$ and $D$ there have been 
functions of a basic unitary operator. Though this formulation includes all 
chiral operators discussed so far as special cases \cite{ke02}, relying on 
the mentioned unitary operator introduces unnecessary restrictions and forms
an obstacle for a more thorough investigation of the indicated symmetries. 

To investigate CP, T and CPT symmetries in a general way, we here first 
analyze the possible properties of the chiral projections starting from the 
Dirac operator and imposing only minimal conditions. We find that due to a 
contribution which inevitably comes with opposite sign in $G$ and $\bG$ one 
generally gets $\bG\ne G$. Furthermore, since the overall sign of the 
respective contribution remains open, it becomes obvious that in the 
construction of the chiral projections one is confronted with two distinct 
possibilities, of which one must be chosen to describe physics. 

We next show that CP transformations as well as T transformations 
interchange the r\^oles of $G$ and $\bG$. This together with the fact that 
one generally has $\bG\ne G$ then is seen to constitute the origin of the 
symmetry violations. With respect to the need of choosing one of the 
mentioned two possibilities in the construction the interchange under CP 
and under T transformations means to violate the original choice. On the 
other hand, CPT symmetry is seen to be generally there and not to be 
affected by $\bG\ne G$.

Finally, considering correlation functions for any value of the index, we
point out that the symmetry violation effects enter them via the bases 
involved. To get quantitative hold of such effects we note that if the 
related interchange of $G$ and $\bG$ would be supplemented by a change of 
the respective sign in the construction one would get symmetry. 
Thus the effect of the violation turns out to be given by the difference of 
the results for the two sign choices and is seen to become manifest in 
entirely different subsets of bases contributing to the correlation functions.

In Section 2 we introduce basic relations and analyze the possibilities
for the chiral projections. In Section 3 we derive the properties for CP, T 
and CPT transformations. In Section 4 we consider the effects caused in 
correlation functions. Section 5 contains our conclusions.

\section{Operator properties}\se

\subsection{Basic relations}

Introducing chiral projections $P_{\pm}$ and $\bP_{\pm}$ with
$P_++P_-=\bP_++\bP_-=\Id$ and requiring that they satisfy
\be
\bP_{\pm}D=DP_{\mp}
\label{DP}
\ee 
we get the decomposition of the Dirac operator into Weyl operators
\be
D=\bP_+DP_-+\bP_-DP_+.
\ee

Considering $P_-$ and $\bP_+$ in the following, we write them in the form 
\be
P_-=\h(\Id-\ga G),\quad\qquad\bP_+=\h(\Id+\bG\ga),
\label{GaG}
\ee
which because of $P_-\dg=P_-=P_-^2$ and $\bP_+\dg=\bP_+=P_+^2$ implies 
unitarity and $\ga$-Hermiticity,
\be
G\1=G\dg=\ga G\ga,\qquad \bG\1=\bG\dg=\ga\bG\ga.
\label{UN}
\ee
According to \re{DP} the operators $G$ and $\bG$ are subject to
\be
D+\bG D\dg G=0.
\label{DG}
\ee

\subsection{Spectral representations}

We consider a finite lattice and require the Dirac operator to be normal, 
$[D\dg,D]=0$, and $\ga$-Hermitian, $D\dg=\ga D\ga$. It then has the spectral
representation
\be
D=\sum_j\hat{\la}_jP_j +\sum_{k}(\la_kP_k\mo{I}+\la_k^*P_k\mo{II}),
\label{specd}
\ee
where the eigenvalues are all different and satisfy $\mbox{Im}\,\hat{\la}_j=0$ 
and $\mbox{Im}\,\la_k>0$. For the orthogonal projections the relations 
$\ga P_j=P_j\ga$ and $\ga P_k\mo{I}=P_k\mo{II}\ga$ hold and we have
\be
\Id=\sum_jP_j+\sum_k(P_k\mo{I}+P_k\mo{II}),
\label{ID}
\ee
where we associate $P_0$ to $\hat{\la}_0=0$, i.e.~$j=0$ to the zero modes of 
$D$.

We require that $G$ and $\bG$ are functions of $D$, which means that their 
eigenvalues are functions of those of $D$. Using this and imposing 
\re{DG} we obtain for $G$ and $\bG$ the spectral representations
\ba
G=\eta_0P_0-\sum_{j\ne0}\eta_jP_j+\sum_k\big(\e^{i\vp_k}P_k
\mo{I}+\e^{-i\vp_k} P_k\mo{II}\big),\nonumber\\\bG=\bar{\eta}_0P_0+
\sum_{j\ne0}\eta_jP_j+\sum_k\big(\e^{i\bar{\vp}_k}P_k
\mo{I} +\e^{-i\bar{\vp}_k}P_k\mo{II}\big),\,
\label{GaG1}
\ea
with $\eta_j$ and $\bar{\eta}_0$ taking the values $\pm 1$ and 
phases $\vp_k$, $\bar{\vp}_k$ being subject to 
\be
\e^{i(\vp_k+\bar{\vp}_k-2\alpha_k)}=-1\qquad\mbox{ where }\qquad
\e^{i\alpha_k}=\la_k/|\la_k|,\qquad 0<\alpha_k<\pi.
\label{PH}
\ee
At this point we make the important observation that because of the
opposite signs of the $j$-sums in \re{GaG1} one generally has $\bG\ne G$.

\subsection{Chiral features}

Since $\ga P_j=P_j\ga$ we get the decomposition $P_j=P_j^++P_j^-$ with 
$\ga P_j^{\pm}=P_j^{\pm}\ga=\pm P_j^{\pm}$. Furthermore $\ga P_k\mo{I}=
P_k\mo{II}\ga$ implies $\Tr(\ga P_k\mo{I})=\Tr(\ga P_k\mo{II})=0$. For 
$N_j^{\pm}=\Tr\,P_j^{\pm}$ then according to $\Tr(\ga\Id)=0$ and \re{ID} 
\be
\sum_j(N_j^+-N_j^-)=0
\label{SU}
\ee
follows. The index of $D$ is given by $I=N_0^{+}-N_0^{-}$. For the numbers 
of the Weyl and anti-Weyl degrees of freedom $N=\Tr\,P_-$ and $\bar{N}=
\Tr\,\bP_+$ we thus obtain 
\be
N=d+\h(-\eta_0I+K),\qquad\bar{N}=d+\h(\bar{\eta}_0I+K),
\label{NbN}
\ee
where $d=\h\,\Tr\,\Id$ and 
\be 
K=\sum_{j\ne0}\eta_j(N_j^+-N_j^-).
\label{KN}
\ee

These relations allow us to discuss \re{GaG1} in more detail. Firstly we
see that \re{NbN} gives $\bar{N}-N=\h(\bar{\eta}_0+\eta_0)I$. Therefore in 
order to have $\bar{N}-N=I$ we must put $\bar{\eta}_0=\eta_0=1$. Next
we note that $\bar{N}+N=2d+K$. If there is only one term in the $j$-sums of 
\re{GaG1}, as has been the case for all operators discussed so far 
\cite{ke02}, using \re{SU} we obtain $K=-\eta_1I$. For $I=0$
this quite reasonably implies $\bar{N}+N=2d$. In the more general case
admitted here putting $\eta_j=\eta$ for $j\ne0$ we get $K=-\eta I$ and
thus also $\bar{N}+N=2d $ for $I=0$. Finally we see that to minimize the
differences between $G$ and $\bG$ the $k$-sums in \re{GaG1} can readily be 
made equal by requiring $\bar{\vp}=\vp$. 
We thus arrive at the form
\ba
G=P_0-\eta\sum_{j\ne0}P_j+\sum_k\big(\e^{i\vp_k}P_k
\mo{I}+\e^{-i\vp_k} P_k\mo{II}\big),\nonumber\\\bG=P_0+\eta
\sum_{j\ne0}P_j+\sum_k\big(\e^{i\vp_k}P_k
\mo{I} +\e^{-i\vp_k}P_k\mo{II}\big),\,
\label{GaG2}
\ea
and remain with the two possible choices $\eta=1$ or $\eta=-1$. The
important observation here is that to describe physics we are forced to 
decide for one of such choices.

\section{Transformation properties}\se

\subsection{CP transformations}

With the charge conjugation matrix
$C$ and with $R^{\cal P}_{n'n}=
\delta^4_{n'n^{\cal\!P}}$, $U_{4n}^{\cal CP}=U_{4n^{\cal\!P}}^*$ and 
$U_{kn}^{\cal CP}=U_{k,n^{\cal\!P}-\hat{k}}^{\rm T}$ for $k=1,2,3$, where 
$n^{\cal P}=(-\vec{n},n_4)$, we have for $D$ the CP transformation
\be
D(U^{\cal CP})=W^{\cal CP}D^{\rm T}(U)W^{{\cal CP}\dag},\qquad W^{\cal CP}=
R^{\cal P}\gamma_4C\dg,
\label{WW}
\ee
in which T denotes transposition and where $W^{{\cal CP}\dag}=
W^{{\cal CP}-1}$ holds.\footnote{With $\gamma_{\mu}\dg=
  \gamma_{\mu}$ and $\gamma_{\mu}^{\rm T}=(-1)^{\mu}\gamma_{\mu}$ for 
  $\mu=1,\ldots,4$ and $\ga=\gamma_1\gamma_2\gamma_3\gamma_4$ 
we have $\ga^{\rm T}=\ga$ and get $C^{\rm T}=-C=C\dg=C\1$ and 
  $[\ga,C]=0$ for $C=\gamma_2\gamma_4$.}
 Because $G$ and $\bG$ are functions of $D$ they inherit its
transformation properties so that
\be
G(U^{\cal CP})=W^{\cal CP} G^{\rm T}(U)W^{{\cal CP}\dag},\qquad 
\bG(U^{\cal CP})=W^{\cal CP}\bG^{\rm T}(U)W^{{\cal CP}\dag}.
\label{GG}
\ee

Using \re{GG} it becomes obvious that the forms \re{GaG},
\be
P_-(U)=\h\big(\Id-\ga G(U)\big),\quad\qquad\bP_+(U)=\h\big(\Id+\bG(U)\ga\big),
\label{PR0}
\ee
because of $\{\ga,W^{\cal CP}\}=0$ transform to
\be
P_-^{\cal CP}(U^{\cal CP})=\h\big(\Id-\ga\bG(U^{\cal CP})\big),\qquad
\bP_+^{\cal CP}(U^{\cal CP})=\h\big(\Id+G(U^{\cal CP})\ga\big),
\label{PRC}
\ee
with $P_-^{\cal CP}(U^{\cal CP})=W^{\cal CP}\bP_+^{\rm T}(U)
W^{{\cal CP}\dag}$ and $\bP_+^{\cal CP}(U^{\cal CP})=W^{\cal CP}P_-^{\rm T}(U)
W^{{\cal CP}\dag}$. Insertion into $I=\bar{N}-N=\Tr\,\bP_+-\Tr\,P_-$ shows
that $I^{\cal CP}=-I$. 

The result \re{PRC} obviously differs from the untransformed relation 
\re{PR0} by an interchange of $G$ and $\bG$. This together with fact that, 
due to the opposite signs of the $j$-sums in \re{GaG1}, one generally has 
$\bG\ne G$ means violation of the symmetry.

With respect to the need that to describe physics one has to choose a 
definite value of $\eta$ in \re{GaG2}, the interchange of $G$ and $\bG$ 
is seen to cause a change violating the original choice of $\eta$.

\subsection{T transformations}

With $R^{\cal T}_{n'n}=\delta^4_{n'n^{\cal\!T}}$, $U_{4n}^{\cal T}=
U_{4,n^{\cal\!T}-\hat{4\,}}^{\rm T}$ and $U_{kn}^{\cal T}=U_{kn^{\cal\!T}}^*$ 
for $k=1,2,3$, where $n^{\cal T}=(\vec{n},-n_4)$, one has for $D$ the
T transformation
\be
D(U^{\cal T})=W^{\cal T}D^{\rm T}(U)W^{{\cal T}\dag},\qquad W^{\cal T}=
R^{\cal T}\ga C\gamma_4,
\label{WWt}
\ee
where $W^{{\cal T}\dag}=W^{{\cal T}-1}$. Therefore we can proceed quite 
analogously to the CP case using 
\be
G(U^{\cal T})=W^{\cal T} G^{\rm T}(U)W^{{\cal T}\dag},\qquad 
\bG(U^{\cal T})=W^{\cal T}\bG^{\rm T}(U)W^{{\cal T}\dag}.
\ee
Because of $\{\ga,W^{\cal T}\}=0$ this leads to 
\be
P_-^{\cal T}(U^{\cal T})=\h\big(\Id-\ga\bG(U^{\cal T})\big),\qquad
\bP_+^{\cal T}(U^{\cal T})=\h\big(\Id+G(U^{\cal T})\ga\big),
\label{PRCt}
\ee
with $P_-^{\cal T}(U^{\cal T})=W^{\cal T}\bP_+^{\rm T}(U)W^{{\cal T}\dag}$,
$\;\bP_+^{\cal T}(U^{\cal T})=W^{\cal T} P_-^{\rm T}(U)W^{{\cal T}\dag}$
and $I^{\cal T}=-I$. 

The result \re{PRCt} obviously differs from the untranformed relation 
\re{PR0} by an interchange of $G$ and $\bG$. Thus again the general mechanism
of symmetry violation takes place which has been described above in the
CP case.

\subsection{CPT transformations}

With $R^{\cal CPT}_{n'n}=\delta^4_{n',-n}$ and $U_{\mu n}^{\cal CPT}=
U_{\mu,-n-\hat{\mu}}\dg$ we get for $D$ the CPT transformation
\be
D(U^{\cal CPT})=W^{\cal CPT}D(U)W^{{\cal CPT}\dag}=R^{\cal CPT}D\dg(U)
R^{\cal CPT},\qquad W^{\cal CPT}=R^{\cal CPT}\ga,
\label{WWs}
\ee
and the analogous relations for $G$ and $\bG$. Using them we obtain
\be
P_-^{\cal CPT}(U^{\cal CPT})=\h\big(\Id-\ga G(U^{\cal CPT})\big),\qquad
\bP_+^{\cal CPT}(U^{\cal CPT})=\h\big(\Id+\bG(U^{\cal CPT})\ga\big).
\label{PRCs}
\ee
with $P_-^{\cal CPT}(U^{\cal CPT})=W^{\cal CPT}P_-(U)W^{{\cal CPT}
\dag}$, $\;\bP_+^{\cal CPT}(U^{\cal CPT})=W^{\cal CPT}\bP_+(U)W^{{\cal 
CPT}\dag}$, $I^{\cal CPT}=I$. 

Obviously \re{PRCs} has the same form as \re{PR0} which means that there
is CPT symmetry. It is seen that this symmetry is not affected by $\bG\ne G$.

\section{Correlation functions}\se

\subsection{Basic relations}

We consider basic fermionic correlation functions in a form which also 
applies in the presence of zero modes and for any value of the index. 
Integrating out the Grassmann variables we have
\be
\langle\psi_{\sigma_{r+1}}\ldots\psi_{\sigma_N}\bar{\psi}_{\bar{\sigma}_{r+1}}
\ldots\bar{\psi}_{\bar{\sigma}_{\bar{N}}}\rangle_{\f}
=\frac{1}{r!}\sum_{\bar{\sigma}_1\ldots\bar{\sigma}_r}\sum_{\sigma_1,\ldots,
\sigma_r}\bar{\Upsilon}_{\bar{\sigma}_1\ldots\bar{\sigma}_{\bar{N}}}^*
\Upsilon_{\sigma_1\ldots\sigma_N}D_{\bar{\sigma}_1\sigma_1}\ldots
D_{\bar{\sigma}_r\sigma_r}
\label{COR}
\ee
with the alternating multilinear forms
\be
\Upsilon_{\sigma_1\ldots\sigma_N}=\sum_{i_1,\ldots,i_N=1}^N\epsilon_{i_1, 
\ldots,i_N}u_{\sigma_{1}i_{1}}\ldots u_{\sigma_Ni_N},\;\; \bar{\Upsilon}_{
\bar{\sigma}_1\ldots{\bar{\sigma}_{\bar{N}}}}=\sum_{j_1,\ldots,
j_{\bar{N}}=1}^{\bar{N}}\epsilon_{j_1,\ldots,j_{\bar{N}}}\bar{u}_{\bar{
\sigma}_{1}j_{1}}\ldots\bar{u}_{\bar{\sigma}_{\bar{N}}j_{\bar{N}}}.
\label{FO}
\ee
The bases $\bu_{\sigma'j}$ and $u_{\sigma i}$ in \re{FO} satisfy
\be
P_-=uu\dg,\quad u\dg u=\Id_{\rm w},\qquad\qquad\bP_+=\bu\bu\dg,\quad\bu\dg\bu
=\Id_{\rm\bw}.
\label{uu}
\ee 
Since by \re{uu} the bases are only fixed up to unitary transformations, 
$u^{[S]}=uS$, $\bu^{[\bar{S}]}=\bu\bar{S}$, we impose the condition 
\be
{\det}_{\rm w}S\cdot{\det}_{\rm\bw}\bar{S}\dg=1
\label{UNI}
\ee
on such transformations in order that general correlation functions remain 
invariant.

\subsection{Transformations}

We now address the case of CP transformations, noting that analogous 
relations hold for T transformations. With conditions \re{uu} and \re{UNI} 
being satisfied by $u$, $\bu$, $S$, $\bS$ as well as by $u^{\cal CP}$, 
$\bu^{\cal CP}$, $S^{\cal CP}$, $\bS^{\cal CP}$, the (equivalence classes 
of pairs of) bases transform as
\be
u^{\cal CP}S^{\cal CP}=W^{\cal CP}\bu^*\bS^*S_{\zeta},\quad\bu^{\cal CP}
\bS^{\cal CP}=W^{\cal CP}u^*S^*\bS_{\zeta},
\label{tCP}
\ee
where the additional unitary operators $S_{\zeta}$ and $\bS_{\zeta}$ have
been introduced for full generality. 
Inserting \re{tCP} into \re{COR} gives for the correlation functions
\ba
\langle\psi_{\sigma_1'}^{\cal CP}\ldots\psi_{\sigma_R'}^{\cal CP}\bar{\psi}_{
\bar{\sigma}_1'}^{\cal CP}\ldots\bar{\psi}_{\bar{\sigma}_{\bar{R}}'}^{\cal CP}
\rangle_{\f}^{\cal CP}=\nonumber\hspace{90mm}\\\e^{i\vartheta_{\cal CP}}
\sum_{\sigma_1,\ldots,\sigma_R}\sum_{\bar{\sigma}_1,\ldots,
\bar{\sigma}_{\bar{R}}}W^{{\cal CP}\dag}_{\bar{\sigma}_1\bar{\sigma}_1'}\ldots
W^{{\cal CP}\dag}_{\bar{\sigma}_{\bar{R}}\bar{\sigma}_{\bar{R}}'}
\quad\langle\psi_{\bar{\sigma}_1}\ldots\psi_{\bar{\sigma}_{\bar{R}}}
\bar{\psi}_{\sigma_1}\ldots\bar{\psi}_{\sigma_R}\rangle_{\f}\,
W^{\cal CP}_{\sigma_1'\sigma_1}\ldots W^{\cal CP}_{\sigma_R'\sigma_R},
\label{CPC}
\ea
where $\e^{i\vartheta_{\cal CP}}={\det}_{\rm\bw}S_{\zeta}\cdot{\det}_{\rm w}
\bar{S}_{\zeta}\dg$. Since repetition of the transformation must lead back, 
$S_{\zeta}$ and $\bar{S}_{\zeta}$ are restricted to choices for which 
$\vartheta_{\cal CP}$ is a universal constant. Accordingly the factor 
$\e^{i\vartheta_{\cal CP}}$ gets irrelevant in full correlation functions 
and, without restricting generality, we can put $\vartheta_{\cal CP}=0$.

Though \re{CPC} then superficially looks ``CP covariant'', it is affected by 
the missing CP symmetry of the chiral projections. Indeed, while the pair
$u$, $\bu$ is related to one choice of $\eta$ in \re{GaG2},  the pair
$u^{\cal CP}$, $\bu^{\cal CP}$ is related to the other one.

\subsection{Symmetry-violation effects}

To get quantitative hold of the symmetry violations we note that for
CP transformations as well as for T transformations an additional change of 
the value of $\eta$ in \re{GaG2} would lead to symmetry. Thus the 
symmetry-violation effect is given by the difference of the results for the 
two choices of $\eta$. 

To study this in detail we note that using \re{GaG2} we obtain
\be
P_-=P_0^-+\sum_{j\ne0}P_j^{\pm}+\sum_kP_k^{\rm X},\qquad
\bP_+=P_0^++\sum_{j\ne0}P_j^{\pm}+\sum_k\bP_k^{\rm X}\qquad\mbox{ for }
\eta=\pm 1,
\label{PbPf}
\ee
where $P_k^{\rm X}=f_k(0)-\ga f_k(\vp_k)$, $\;\bP_k^{\rm X}=f_k(0)+
f_k(\vp_k)\ga$ with $f_k(\vp)=\h(\e^{i\vp}P_k\mo{I}+\e^{-i\vp}P_k\mo{II})$. 
\hspace{0mm}From the representation \re{PbPf} 
it becomes obvious that for the two choices of $\eta$ either 
$\;\sum_{j\ne0}P_j^+\;$ or $\;\sum_{j\ne0}P_j^-\;$ is involved in the
chiral projections.

The crucial point now is that this implies the occurrence of the entirely 
different contributions $\;\sum_{j\ne0}P_j^+=\sum_{l=1}^{L^+}u_l^+u_l^{+\dag}
\;$ and $\;\sum_{j\ne0}P_j^-=\sum_{l=1}^{L^-}u_l^-u_l^{-\dag}\;$ to \re{uu}
$\;$ (where since $L^--L^+=I$ even the numbers of terms can differ). In the 
multilinear forms \re{FO} then the subsets $u_l^+$ and $u_l^-$ of bases, 
being related to entirely different projections, clearly lead to different 
results. This in turn causes differences in the correlation functions 
which give the effects of the symmetry violations.

There remains obviously the question which one of the two choices for $\eta$
is the appropriate one for the description of physics. However, at present 
no theoretical principle is in sight to decide about this.

\section{Conclusions}

We have investigated CP, T and CPT symmetries in a general way, imposing only 
minimal conditions,  namely normality and $\ga$-Hermiticity of the 
Dirac operator and that it has a general decomposition into Weyl operators.

We first have analyzed the possible properties of the chiral projections 
starting from the Dirac operator. It has turned out that due to a contribution
in the spectral representations which inevitably comes with opposite sign in 
the operators $G$ and $\bG$, which enter the chiral projections, one generally gets $\bG\ne G$. Furthermore, because the overall 
sign of the respective contribution remains open, it has become obvious that 
in the construction of the chiral projections one is confronted with two 
distinct possibilities, of which one must be chosen to describe physics. 

We next have shown that CP transformations as well as T transformations 
cause an interchange of the r\^oles of $G$ and $\bG$. This together with the 
observation that one generally gets $\bG\ne G$ has been seen to constitute 
the origin of the symmetry violations. With respect to the need of choosing 
one of the mentioned two possibilities in the construction it has become 
obvious that the interchange of $G$ and $\bG$ under CP and under T 
transformations means violation of the original choice. On the other hand, 
CPT symmmetry has been seen to be generally there and not to be affected by 
$\bG\ne G$. 

Finally, using a form of the correlation functions which applies also in 
the presence of zero modes and for any value of the index, we have pointed 
out that the symmetry-violation effects enter them via the bases involved. 
To get quantitative hold of such effects we have noted that if the related 
interchange of $G$ and $\bG$ would be supplemented by a change of the 
respective sign in the construction one would get symmetry. Thus the effects 
of the violations have turned out to be given by the difference of the results 
for the two choices in the construction of the chiral projections. This has 
been seen to become manifest in entirely different subsets of bases appearing 
in the correlation functions.

\section*{Acknowledgement}

I wish to thank Michael M\"uller-Preussker and his group for their kind
hospitality.

\end{document}